\documentclass[a4paper]{jpconf}
\usepackage{graphicx}
\usepackage{hyperref}
\usepackage{amsmath}
\usepackage{amsthm}
\usepackage{amssymb}
\def\tr{\textrm{tr}\,}
\begin{document}
\title{From Feynman graphs to Witten diagrams}

\author{Domingo Gallegos, Umut G\"ursoy, Natale Zinnato}

\address{Institute for Theoretical Physics, Utrecht University,
Leuvenlaan 4, 3584 CE Utrecht, The Netherlands}

\ead{a.d.gallegospazos@uu.nl, u.gursoy@uu.nl, n.zinnato@uu.nl}

\begin{abstract}
We investigate the possibility of generalizing Gopakumar's microscopic derivation of Witten diagrams in large N free quantum field theory \cite{Gopakumar:2003ns} to interacting theories. For simplicity we consider a massless, matrix valued real scalar field with $\Phi^h$ interaction in d-dimensions. Using Schwinger's proper time formulation and organizing the sum over Feynman graphs by the number of loops $\ell$, we show that the two-point function can be expressed as a sum over boundary-to-boundary propagators of bulk scalars in $AdS_{d+1}$ with mass determined by $\ell$. 

\end{abstract}

\section{Introduction}
\noindent\\
Gauge-string correspondence \cite{Maldacena:1997re,Witten:1998qj,Gubser:1998bc}, despite all the successful checks it went through and the plethora of work that applied it to study, again successfully, both strongly interacting QFTs and quantum gravity, lacks a satisfactory microscopic derivation and 
a deeper understanding thereof. A basic question is how to reformulate holographic QFT correlation functions, in particular Feynman graphs, such that emergence of gravitational dynamics becomes manifest, for example how the gravitational field propagators in the dual curved space-time arise from field theory amplitudes. Another intriguing question is, how to determine which QFTs are holographic, which are not. Furthermore, given a holographic QFT and assuming a limit where the dual geometry is well defined, is there an algorithm to determine this dual background directly from the QFT correlators? 

Among all the different approaches that have been suggested in the literature, entanglement entropy \cite{Ryu:2006bv}, geometrization of RG flows \cite{Heemskerk:2010hk,deBoer:1999tgo,Lee:2013dln}, bulk reconstruction \cite{Harlow:2018fse}, quantum error correction \cite{Almheiri:2014lwa}, tensor networks \cite{Hayden:2016cfa}, etc., there is one which stands out as the most elementary: deriving dual gravity propagators directly from the QFT Feynman graphs. As far as we know, for field theories in $d>2$ this approach was first proposed by R. Gopakumar in the case of free field theory \cite{Gopakumar:2003ns}. Gopakumar considered  matrix valued free field theories and studied n-point function of composite single-trace operators in Schwinger's proper time formulation \cite{Schwinger:1951nm}. In the particular case of the three-point function the author considered a change of variables involving the moduli (Feynman parameter's of a given graph) that is called the {\em star-triangle duality}\footnote{Earlier work relating matrix quantum mechanics and 2D non-critical string theory \cite{Kazakov:1985ea} relies on a similar type of duality.}. The name derives from an analogous relation that involves electric circuits\footnote{See for example, \cite{Lam:1969xk} for a precise account of the map between Feynman graphs and electric circuits.} which relates the total effective impedance of a triangle shaped electric circuit to that of a tri-star circuit, see fig. \ref{fig2}. In Schwinger's formulation the total proper time of the graph is in one-to-one correspondence with the holographic direction of the dual gravity theory\footnote{See also \cite{Gopakumar:2004qb,Gopakumar:2005fx,Gopakumar:2004ys}.} \cite{Gopakumar:2003ns} and the star-triangle relation becomes a clear manifestation of the gauge-string duality, or open-closed duality in string theory where the gauge theory 3-point function is represented by the triangle and the corresponding Witten diagram \cite{Witten:1998qj} in the dual theory is represented by the tri-star. See \cite{Aharony:2020omh} for a more recent work, based on a different approach, that also derives dual gravity theory directly from field theory, in the case of vector models \cite{Gubser:2002tv}.   

In this note we suggest that a generalization of the star-triangle type duality of Feynman graphs into interacting field theories might be a fundamental manifestation of the gauge-string duality and a key to generalize it beyond the known specific cases\footnote{e.g. based on D-brane descriptions \cite{Aharony:1999ti}, lower dimensional examples \cite{Klebanov:1991qa,Maldacena:2016hyu} and vector models \cite{Aharony:2020omh,Gubser:2002tv}.}. In particular, we generalize Gopakumar's derivation of Witten's diagrams from free field theory to interacting theories. As a prototype, we take a real, massless N$\times$N matrix-valued scalar field $\Phi$ with $\Phi^h$ interaction, for integer $h>2$, in  d-dimensions, and consider the two-point function $\langle \Phi(x_1) \Phi(x_2) \rangle $. The two-point function in the large N limit is given by Feynman graphs summed over the number of independent quantum loops $\ell$ which, in the large N limit, can further be classified in terms of 2D Euclidean Riemann surfaces embedded in d-dimensions. We show that each term in the sum over $\ell$ can be mapped onto a boundary-to-boundary propagator of a scalar field with mass $m$ related to $\ell$, in $d+1$ dimensional AdS space. This provides a dual ``closed string'' picture of the two-point function in terms of a {\em generalized Witten diagram} given by sum over AdS Witten diagrams. Even though we perform our calculations in a simple scalar ungauged theory, we will be assuming that our findings generalize to theories like ${\cal N}=4$ super-Yang-Mills without conceptual difficulties. 

In the next section we review Gopakumar's construction in free field theory. In section 3 we first review Schwinger's proper time formulation for interacting field theories and then generalize Gopakumar's computation to finite coupling. We end in section 4 by discussing open issues, generalization to higher point functions and an overlook.

\section{Open-closed and star-triangle}
\begin{figure}
\begin{center}
\includegraphics[width=15cm]{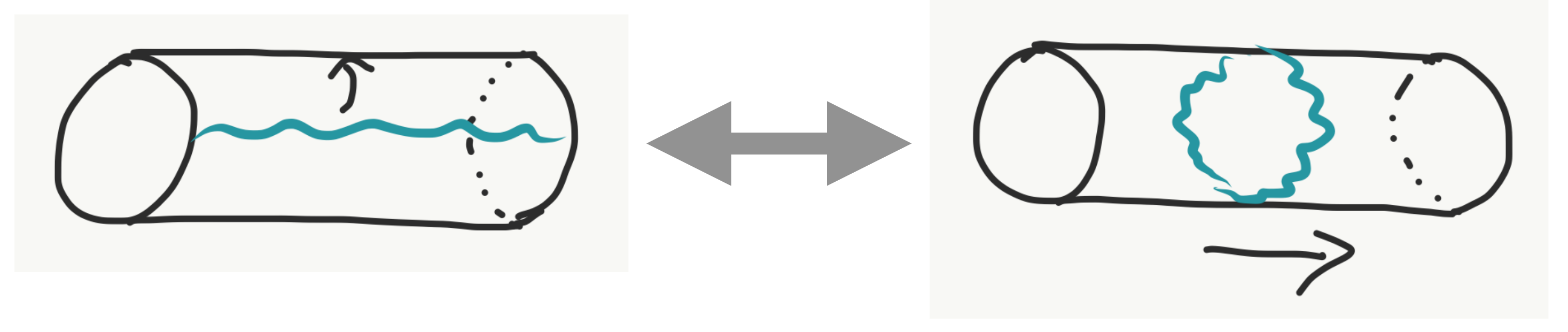}
\end{center}
\caption{\label{fig1} Equivalence between one-loop diagram of an open string and tree level propagation of a closed string.}
\end{figure}

\noindent\\
The AdS/CFT correspondence originates from an equivalence between open and closed string  descriptions of a set of D3 branes in IIB string theory \cite{Maldacena:1997re}. Loosely speaking, and in the simplest case, this can be understood geometrically as in fig. \ref{fig1} which depicts an equivalence between one-loop partition function of open strings in d-dimensions and propagation of a closed string in $d+1$ dimensions \cite{Polyakov:1987ez}.  
In the low energy limit where the massive string states decouple, open strings on N coincident D3 branes are effectively described by 4D $U(N)$ Yang-Mills gauge theory theory. On the other hand the closed string in fig. \ref{fig1} turns out to propagate in the AdS$_5\times S^5$ geometry which is generated by the backreaction of the brane system. More precisely, 
the n-point function of gauge invariant operators in the Yang-Mills theory is given 
in terms of the closed string world-sheet path integral  
\begin{equation}
\label{eq:oc}
\langle {\cal O}_1 (k_1) \cdots  {\cal O}_n (k_n) \rangle_g = \int_{M_{g,n}} \langle {\cal V}_1(k_1,z_1) \cdots  {\cal V}_n(k_n,z_n)\rangle_{w.s.}\, ,
\end{equation}
where the subscript $g$ on the RHS denotes the genus-g contribution to the Feynman graphs and ${\cal V}$s are the closed string vertex operators which correspond to gauge theory operators on the LHS. The integral is over the moduli of Riemann surfaces with genus g and n punctures. 

To demonstrate this equivalence at the level of Feynman graphs, one must show, how the holes on the open string (gauge theory) side are glued together and generate closed string world-sheets with n-punctures. This mechanism was first proposed by 't Hooft \cite{tHooft:1973alw}
in the double scaling limit, 
\begin{equation}\label{thooft}
 g_{YM} \to 0\, , \qquad N\to\infty\, , \qquad g_{YM}^2 N = \lambda \, ,
\end{equation}
where $g_{YM}$ is the Yang-Mills coupling constant. Emergence of a dual description in this limit can be made explicit in 2D string theory, where the quantum mechanics of $N\times N$ hermitean matrices become dual to 2D non-critical string theory, see for example \cite{Klebanov:1991qa}. 

A strong indication that the same ``gluing'' happens in higher dimensional {\em free} field theories was noted in \cite{Gopakumar:2003ns} utilising the proper time formulation of 
n-point functions, which we review below. 

Schwinger's proper time formulation makes the point-like feature 
of QFT manifest. In particular, correlators of a quantum field are
represented by propagation of quantum mechanical particle in proper 
time $\tau$ embedded in space-time as world-line $x^\mu(\tau)$. 
To see this one exponentiates the denominator in the two-point function 
\begin{equation}\label{sch1}
  \langle \Phi(x_1) \Phi(x_2) \rangle = i \int d^dk \frac{e^{i k(x_1-x_2)}}{k^2 + m^2 - i\epsilon} = \int_0^\infty d\tau \langle x_1 | e^{-i \tau (-\partial^2 + m^2)} | x_2 \rangle   \, .
\end{equation}
The RHS is nothing else but the path integral of a particle propagating in $\tau$ with hamiltonian $H_{pp} = k^2 + m^2$. The integral over $\tau$ is moduli --- a consequence of the reparametization invariance of the worldsheet \footnote{which can be removed by introducing an auxiallry worldsheet einbein $g^{\tau\tau}$ in the path integral.}. A generalization of this representation to n-point functions in a free field theory involves introduction of vertex operators inside the path integral
\begin{equation}
\label{eq:npf}
    \langle \phi(x_1) \cdots \phi(x_n) \rangle = \int_0^\infty \frac{d\tau}{\tau} \prod_{i=1}^n d\tau_i \langle e^{i k_1 \hat X(\tau_1)}\dots e^{i k_n \hat X(\tau_n)} \rangle_{q.m.} 
\end{equation}
where the RHS is the path integral with the point particle hamiltonian $H_{pp} = k^2 + m^2$. 
The integral is over the moduli of the Feynman graph given by the total proper time for the process and proper times at insertions of the vertex operators. Note the structural similarity between (\ref{eq:oc}) and (\ref{eq:npf}) which already implies the utility of the 
Schwinger's formulation to explore the basic mechanism behind the gauge-string duality. 

\begin{figure}
\begin{center}
\includegraphics[width=15cm]{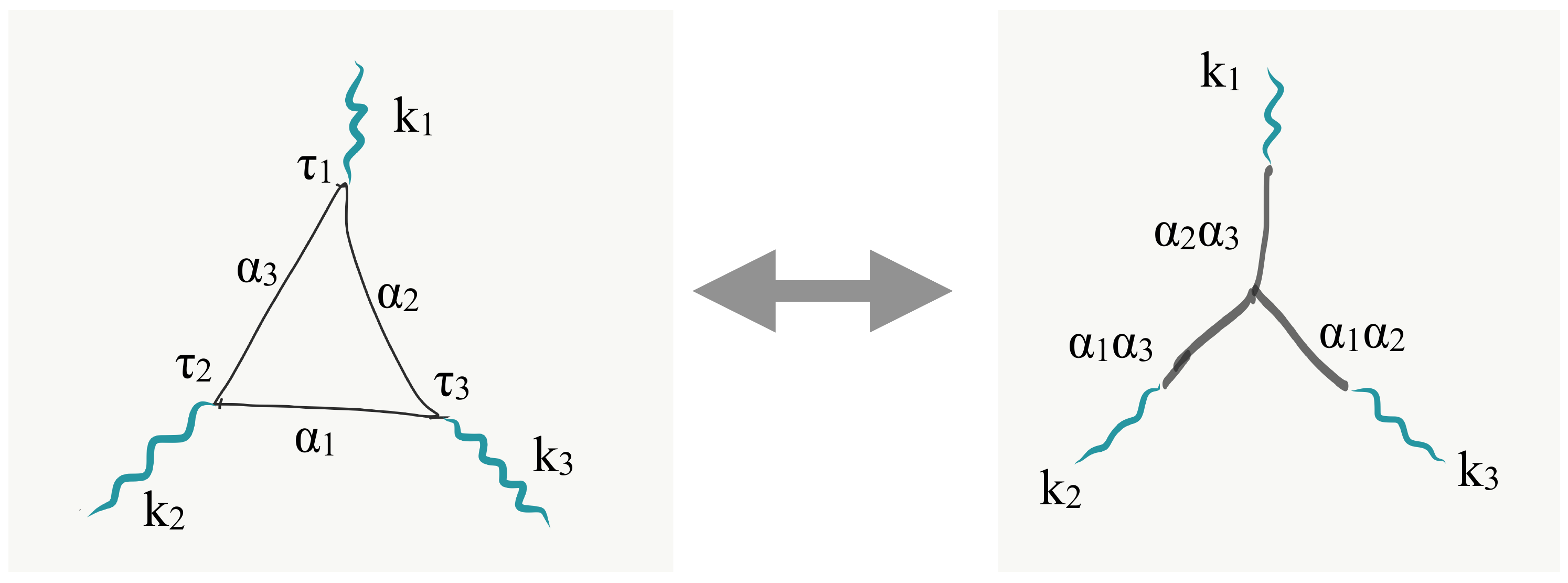}
\end{center}
\caption{\label{fig2} Star-triangle duality in free field theory. LHS shows the only Feynman graph that contributes to the 3-point function of $\tr \Phi^2$ where $k_i$ are external momenta, $\tau_i$ are the moduli and $\alpha_i$ are the Schwinger parameters. RHS shows its equivalent under the duality.}
\end{figure}

As the path integral in (\ref{eq:npf}) is Gaussian for free field theory, one can compute it explicitly \cite{Strassler:1992zr} and express the result solely in terms of moduli integrals. More interestingly, one can find a judicious change of variables of moduli to reformulate the result in terms of propagators of scalar fields in AdS$_{d+1}$ \cite{Gopakumar:2003ns,Gopakumar:2004ys}. Consider ${\cal N}=4$ super-Yang-Mills at large N and in the free limit $\lambda=0$, see (\ref{thooft}). For the purpose of demonstration let us consider the simplest non-trivial case of the 3-point function and the operator $\tr \Phi^2$ where $\Phi$ is one of the 6 scalars in the theory. There is a single diagram that contributes to the connected 3-point function $\langle \tr \Phi^2(k_1) \Phi^2(k_2) \Phi^2(k_3)\rangle $ that is shown on the left figure in fig. \ref{fig2}. 

Introducing a change of variables \cite{Gopakumar:2003ns} $\alpha_i = \epsilon_{ijk} |\tau_j - \tau_k|/\tau$ from the moduli $\tau_i$ to Schwinger parameters  one can rewrite the connected 3-point function as follows 
\begin{equation}
    \Omega(k_1,k_2,k_3) \propto \delta^d(\sum k_i) \int_0^\infty d\tau \int_0^1 \prod_{i=1}^3 d\alpha_i \, \delta(\sum \alpha_i -1)\,  e^{-\tau(k_1^2 \alpha_2\alpha_3 + k_2^2 \alpha_3\alpha_1 + k_3^2 \alpha_1\alpha_2)}
\end{equation}
This is precisely in the form given by product of three propagators with 
dual Schwinger parameters $\alpha_1\alpha_2$ etc. as shown on the RHS of fig. \ref{fig2}. This procedure explicitly achieves the ``gluing'' mentioned above in the sense that the hole on the ``open string side'' i.e. the LHS of fig. \ref{fig2} is closed up on the ``closed string side'' i.e. the RHS of fig. \ref{fig2}. The RHS also resembles the Witten diagram for the 3-point function in AdS and this resemblance can be made precise by another change of variables $\alpha_i = \rho_i/\sum_{j=1}^3 \rho_j$ \cite{Gopakumar:2003ns} and defining the radial coordinate of the AdS space $z_0$ in terms of these Schwinger moduli as 
\begin{equation}
    z_0^2 = 4\tau \left(\sum_{i=1}^3 \rho_i \right)\prod_{i=1}^3 \alpha_i\, .
\end{equation}
This results in the final expression after Fourier transforming to space-time as 
\begin{equation}
    \Omega(x_1,x_2,x_3) \propto  \int_0^\infty \frac{dz_0}{z_0^{d+1}} \int d^dz \prod_{i=1}^3 K_{\Delta_i}(x_i;z,z_0)
\end{equation}
where $K_{\Delta}(x_i,y;z)$ are the boundary-to-bulk propagator for a scalar with mass $m^2 = \Delta(d-\Delta)$, with $\Delta = 2$, corresponding to $\tr \Phi^2$ operator in AdS$_{d+1}$ on the Poincar\'e patch
\begin{equation}
   ds^2 = \frac{1}{z_0^2} \left(dz_0^2 + \eta_{ab}\, dz^a dz^b \right)\, . 
\end{equation}
 This computation can be generalized to an arbitrary string of $\Phi$ fields \cite{Gopakumar:2004qb}, presumably to other ${\cal N}=4$ super-Yang-Mills operators and higher point functions \cite{Gopakumar:2004ys}. 

\section{Generalization to interacting theories}
\noindent\\

The derivation of AdS propagators from Feynman graphs in the free case, presented above, carries over to the interacting QFTs \cite{us} to a large extent. For simplicity, we will discuss two-point functions in a massless matrix valued scalar field $\Phi$ in d-dimensions with an interaction potential $\Phi^h$. The action is
\begin{equation}\label{action00}
\mathcal{S} = \int d^dx\, \text{Tr}\, \left(-\frac{1}{2}(\partial\Phi)^2 +  \mathfrak{g}\Phi^h\right)\, ,
\end{equation}
where $h>2$ is the coordination number of the vertex associated with the interaction term $\Phi^h$ and $\mathfrak{g}$ is the coupling constant analogous to $g_{YM}^2$ in (\ref{thooft}). After the rescaling $\Phi \rightarrow \sqrt{N} \Phi$ one obtains
\begin{equation} \label{action0}
\mathcal{S} = N\int d^dx\, \text{Tr}\, \left(-\frac{1}{2}(\partial\Phi)^2  +\lambda\Phi^h\right)\, ,
\end{equation}
where $\lambda \equiv  N^{(h-2)/2}\mathfrak{g}$ is the 't Hooft coupling. 

We are interested in computing correlation functions of scalar fields in \eqref{action0}. We consider a Feynman diagram $F$ of genus $g_0$, with $I$ internal lines, $V$ vertices and $\ell=I-V+1$ independent loops\footnote{We use the double line notation of 't Hooft implicitly. A ``line'' actually refers to a double line in the discussion below.}. Using Euler's formula, $V-I+f = 2-2g_0$, we can also relate the number of loops $\ell$ to the number of faces $f$ via 
\begin{equation} \label{loopface}
\ell = f-1+2g_0\, .
\end{equation}

Schwinger's proper time formulation can be generalized  \cite{Lam:1969xk} to express the value, $\Omega_F$, of this generic graph $F$ in terms of integrals of Schwinger parameters $a_r$, $r = 1,\cdots I$ introduced for each internal line\footnote{We suppress the matrix indices in what follows.}   (analogous to $\tau$ in (\ref{sch1})): 
\begin{equation} \label{npt}
\Omega_F(\vec k)  =\delta^{(d)}\left(k_1+\dots+k_{n_e}\right)\int_{0}^{\infty}\left( \prod_{r=1}^{I}da_r\right)\, \mathcal{U}(a_r)^{-d/2}\, e^{-P\left(a_r;\vec k\right)} \, .
\end{equation}
This is the expression for the amputated graph in Euclidean time with $n_e$ external momenta which we collectively denote as $\vec k =(k_1,\dots, k_{n_e})$. The non-trivial ingredients here are the {\em Symanzik polynomials} 
\begin{equation}
    \mathcal{U}(a) \equiv\sum_{T_1\in \mathcal{T}_1}\prod^{\ell}_{r\not \in T_1} a_r \, , \qquad \ \
P(a_r;\vec k) \equiv \frac{1}{\mathcal{U}(a_r)}\sum_{T_2\in  \mathcal{T}_2}\left(\prod^{\ell+1}_{r\not \in T_2}a_r\right)\left(\sum_{b\in \mathcal{J}}k_b \right)^2 \, ,
\end{equation}
with $\mathcal{T}_1,\mathcal{T}_2$ being the sets of trees and 2-trees respectively\footnote{If one removes $\ell$ internal lines from $F$ such that there are no loops left, the remaining graph can be shown to be a simply-connected subgraph of $F$, which is called a \textit{tree}, $T_1$. If $\ell+1$ lines are removed then one is left with two disconnected components (trees) with no loops, called  a \textit{2-tree}, $T_2$.},  and $\mathcal{J}$ is one of the two disconnected components of a 2-tree. 

Applying this to the connected two-point function and after a series of change of variables \cite{us} and Fourier transform to position space one arrives at the following compact expression 
\begin{equation}
     \label{Omsol0}
\Omega_F(x, y)= 4^{\Delta-2} \pi^{d/2}\Gamma(\Delta) \frac{\mathcal{V}_F }{|x-y|^{2\Delta}}\,. 
\end{equation}
This has  precisely  the same form as a CFT$_d$ two-point function $\langle {\cal O}_\Delta {\cal O}_\Delta \rangle$ of a conformal field ${\cal O}_\Delta$ with the scaling dimension
\begin{equation}\label{Delta2pt}
\Delta = \left(\frac{d}{2}-\frac{h}{h-2}\right)\ell +\frac{d}{2}+1 \, ,
\end{equation}
where we expressed the variables $I$ and $V$ in terms of $\ell$ using Euler's theorem for genus-0 graphs and an additional relation between $I$, $V$ and the coordination number of the vertex $h$. It is tempting to call this quantity as the ``conformal dimension of the Feynman graph''. Note that it does not depend on the detailed structure of the graph but only on the number of independent loops $\ell$.

Moreover, the entire dependence on the moduli $a_r$ is contained in the overall coefficient $\mathcal{V}_F$ in (\ref{Omsol0}) that is given by
\begin{equation}\label{Vn}
\mathcal{V}_F\equiv\ell\int_0^\infty\left(\prod_{r=1}^{I} da_r\right) \, \delta\left(1-\mathcal{U}_F(a) \right)\mathcal{A}_F(a)^{\Delta-d/2}\, ,\qquad 
\mathcal{A}_F(a)\equiv  \sum_{T_2\in \mathcal{T}_2}\prod^{\ell+1}_{r \not \in T_2} a_r\, .
\end{equation} 
This coefficient depends on the particular Feynman graph through the sets $\mathcal{T}_1,\mathcal{T}_2$ in the Symanzik polynomials specific to a given graph $F$. 

\begin{figure}
\begin{center}
\includegraphics[width=16cm]{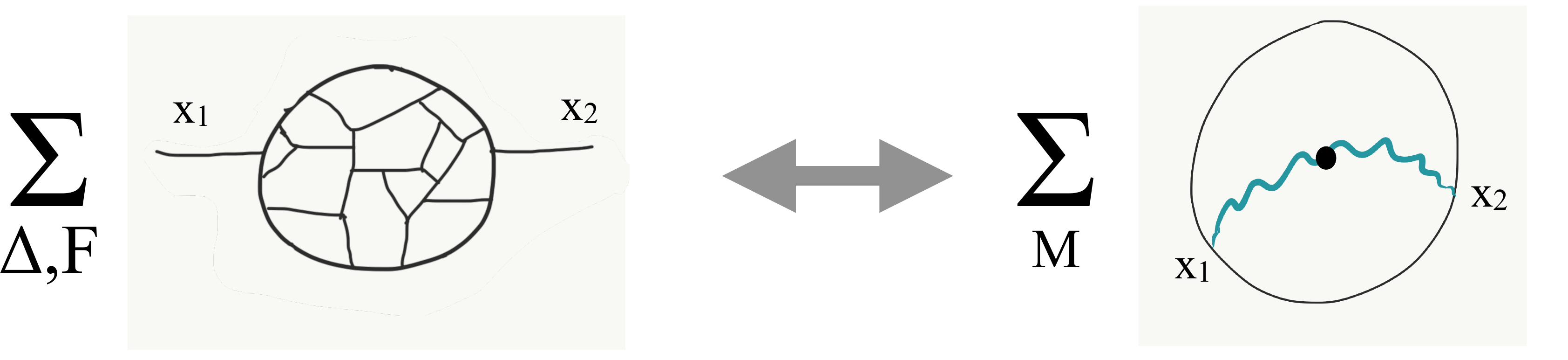}
\end{center}
\caption{\label{fig3} A schematic description of the duality between Feynman graphs and Witten diagrams. On the LHS the sum is over the scale dimension of a graph $\Delta$, equivalently over the number of loops $\ell$, see equation (\ref{Delta2pt}), and over all Feynman graphs with the same $\Delta$. On the RHS the sum is over two-point Witten diagrams that invovle a bulk scalar with mass $M = \Delta (d-\Delta)$. }
\end{figure}

The fact that the contribution to the two-point function is in the CFT form (\ref{Omsol0}) immediately suggests, through the standard AdS/CFT prescription, that the full two-point function can be written as a sum over two-point Witten diagrams as shown in fig. \ref{fig3}. This can indeed be shown explicitly \cite{us} again following a judicious change of variables
similar to section 2. 
The Witten diagram corresponding to the graph F, equation (\ref{Omsol0}) reads\footnote{The proportionality constant involves a UV cut-off, for the more precise expression we refer to \cite{us}.}
\begin{equation} \label{2pfell}
\Omega_F(x_1, x_2) =  \mathcal{V}'_F \lim_{\epsilon\rightarrow 0} \epsilon \int\frac{dz_0 d^dz}{z_0^{1+d}}\, z^{2\epsilon}\,K_{\Delta+\epsilon}(x_1; z, z_0) K_{\Delta+\epsilon}(x_2; z, z_0) \, ,
\end{equation}
where we defined a new coefficient $\mathcal{V}'_F=\mathcal{V}_F\pi^d4^{\Delta}\Gamma(\Delta-\frac{d}{2})$ and introduced the AdS bulk-to-boundary propagator 
\begin{equation} \label{AdSProp0}
K_{\Delta}(x; z, z_0)=\frac{\Gamma(\Delta)}{\pi^{d/2}\Gamma(\Delta-\frac{d}{2})}\frac{z_0^{\Delta}}{\left(z_0^2+(x-z)^2\right)^{\Delta}}\, ,
\end{equation}
of a bulk field with mass $M = \Delta(d-\Delta)$. It is straightforward to check that the limit in (\ref{2pfell}) is finite. 

The full perturbative two-point function in the large N limit is then given by the sum over the number of independent quantum loops $\ell$ of the contributing connected Feynman graphs\footnote{The overall factor of $N$ depends on the operator. For example, the free field scaling for the composite operator $\tr{\Phi^m}$ in $U(N)$ gauge theory, before the rescaling below equation (\ref{action00}), is $N^m$, see e.g. \cite{DHoker:1998vkc}. The expression here is after the rescaling and we consider $m=1$ for simplicity. Generalization to composite operators is straightforward \cite{us}.} 
\begin{align}
\begin{split} \label{fullGsum}
\Omega(x_1,x_2)= \lim_{\epsilon\rightarrow 0} \sum_{\ell=0}^\infty \,\lambda^{2\ell/(h-2)}\,\mathcal{C}_\ell\, \epsilon\int_{AdS}z_0^{2\epsilon} K_{\Delta+\epsilon}(x_1;z,z_0) K_{\Delta+\epsilon}(x_2;z,z_0)\  \, ,
\end{split}
\end{align}
where the coefficient
\begin{equation}\label{density}
    \mathcal{C}_\ell \equiv  \sum_{F\in \mathcal{F}_\ell}\frac{\mathcal{V}'_F}{\sigma_F}
\end{equation}
is given by the sum over all connected Feynman graphs with $\ell$ independent loops and the symmetry factor $\sigma_F$. This is our final expression for the two-point function expressed in terms of AdS propagators. It is a strikingly simple expression where all the complication from distinct Feynman graphs is absorbed in a single coefficient $\mathcal{C}_\ell$. 
\section{Discussion}
\noindent\\

Our final result (\ref{fullGsum}) contains divergences and should be understood as a formal expression. Apart from the standard issue of Borel summability of perturbative expansion in QFT, summands in (\ref{fullGsum}) contain multiple UV and IR divergences contained in the coefficients $\mathcal{V}'_F$. The UV divergences all arise from independent loops in the Feynman graphs, in the limit $a_r\to 0$ of Schwinger parameters. These can be regulated in the standard manner by renormalizing the two-point function. The IR divergences, on the other hand, arise from the other boundary limits $a_r\to \infty$ which can also be regularized by introducing a small mass term in (\ref{action0}). These issues will be detailed in \cite{us}. 

Independent of its holographic interpretation, (\ref{fullGsum}) is interesting in the sense that it corresponds to an alternative form of the K\"all\'en-Lehmann representation of the two-point function in massless QFT. Indeed the sum over $\ell$ can easily be turned into an integral over $\Delta$ which labels excitations in the Hilbert space. Then the density of these states are determined by the coefficient in (\ref{density}). The results we obtained here for the two-point function seem to generalize to higher point-functions barring some technical difficulties that will be addressed in \cite{us}. In case of the three-point function this construction implies an interacting field theory analog of the star-triangle duality, fig. \ref{fig2}, which could be playing a central role in derivation of the gauge-string correspondence for generic holographic QFTs. 

It is also tempting to speculate on the holographic interpretation of this formula. Assuming that the QFT we started with is holographic --- for example ${\cal N}=4$ super-Yang-Mills or its marginal or relevant deformations --- then the RHS of (\ref{fullGsum}) would correspond to a closed string propagator in a curved space-time that is asymptotically AdS$_{d+1}$. In the large N limit we are considering here, this is given by the genus-0 contribution to the world-sheet path integral in (\ref{eq:oc}). Is it then possible to interpret the sum, (\ref{fullGsum}), over $\ell$  as contribution of different string states to the string propagator in this curved spacetime? Is this space-time a solution to $(d+1)$ dimensional non-critical string theory? Is there a saddle point in this sum (which can be expressed as an integral over $\Delta$) that corresponds to the gravity limit analogous to $\lambda\to\infty$ in  ${\cal N}=4$ super-Yang-Mills? Can we then read off the metric in this space-time from the coefficients $\mathcal{C}_\ell$? Would it then be possible to use our approach as an operational definition of the holographic dual to a given QFT e.g. QCD? 
We hope that the approach we outlined in this paper will be instrumental for a deeper understanding of holographic duality and will shed new light on such fundamental questions.

\end{document}